\documentclass[reprint,superscriptaddress,nofootinbib,prd,floatfix]{revtex4}
\pdfoutput=1
\usepackage{graphicx}
\usepackage[dvipsnames]{xcolor}
\usepackage{amsmath}
\allowdisplaybreaks
\usepackage{amssymb}

\usepackage{slashed}
\usepackage[utf8]{inputenc}
\usepackage[colorlinks=true,linkcolor=blue,bookmarksopen,bookmarksnumbered]{hyperref}
\usepackage{color} 

\usepackage[normalem]{ulem}
\usepackage{soul}
\usepackage{units}
\usepackage{rotating}
\usepackage{hhline,multirow,tabularx}
\usepackage{bm}

\usepackage{cleveref} 
\usepackage{nameref} 
\usepackage[export]{adjustbox}
\usepackage{mdframed}
\usepackage{comment}
\usepackage{natbib}
\usepackage{aliascnt} 

\def\bea#1\eea{\begin{align}#1\end{align}}

\usepackage{lipsum}
\usepackage{hyperref}
\begin{document}

\title{Impact parameter dependence of the effect of background field on coupling constant in heavy ion collisions}

\author{Cong Li}
\affiliation{School of Information Engineering, Zhejiang Ocean University, Zhoushan, 316022, Zhejiang, China}


\begin{abstract}

We investigated the impact parameter dependence of the background field's effect on the coupling constant of the $\gamma \gamma \rightarrow l^{+} l^{-} \gamma$ process in heavy-ion collisions. The peripheral electric fields of heavy ions collide, and after photon annihilation into lepton pairs, the subsequent emission of radiation photons will be affected by the background field. Numerical estimates indicate that the dependence of this effect on the impact parameter is significant during this process.

\end{abstract}
\maketitle

\section{Introduction}

In quantum chromodynamics (QCD), the strong interaction coupling constant decreases with increasing energy scale, a phenomenon known as asymptotic freedom. This indicates that the interaction strength between quarks and gluons diminishes at high energy, representing a significant theoretical achievement in particle physics.

The color glass condensate (CGC) is a predicted state of matter in high-energy hadrons, such as protons and heavy ions. As the momentum fraction $x$ decreases, the density of quarks and gluons increases sharply until a critical point is reached, leading to a gluon-saturated state \cite{CGC1,CGC2}. Another state of matter, the quark-gluon plasma (QGP), occurs at extremely high temperatures and densities, where quarks and gluons are no longer confined within hadrons and can move freely. QGP is expected to exist during heavy-ion collisions and in the early universe \cite{QGP1,QGP2}.

However, scattering processes in these complex backgrounds pose theoretical challenges. In contrast to the simplified model of perturbative quantum chromodynamics (pQCD), scattering in QGP and CGC is significantly more complex. Under high density and short distance conditions, asymptotic freedom becomes less applicable, although intuitively it is true. In QGP, intricate high-temperature and high-density interactions undermine simplified theoretical predictions. The gluon saturation in CGC indicates that the coupling constant's behavior in high-density environments may differ significantly from that in low-density ones, challenging the applicability of perturbation theory. \cite{CGC1,CGC2,QGP1,QGP2}

Above all, we want to explore coupling constant corrections in complex backgrounds and quantitatively analyze the influence of these background fields. Understanding these corrections is vital for deepening our knowledge of strong interaction theory and has significant implications for high-energy physics and early universe studies. By clarifying how background fields affect the coupling constant, we aim to uncover new physical phenomena and theoretical challenges in QCD under complex conditions, which is crucial for advancing cutting-edge particle physics research. For instance, the correction of the strong interaction coupling constant in the hadronic background field is crucial for the accuracy of theoretical predictions regarding QCD processes in that environment.

In our previous work \cite{Li3}, we examined corrections to the electromagnetic coupling constant due to background photon fields in heavy-ion ultra-peripheral collisions (UPCs). We selected this process for three reasons: first, the well-studied QED process offers valuable insights applicable to the QCD process, which is also a gauge theory. Second, recent heavy-ion collision experiments have produced vast amounts of data and improved detection technologies, providing strong experimental support for UPC studies \cite{star,star2021,atlas}. Lastly, recent theoretical research has extensively analyzed the structure function of external photons, offering a stable background field essential for these investigations \cite{Li1,Li2,Shao1}.

In this work, we investigate the dependence of the impact parameter in the same scattering process, which is crucial for distinguishing between peripheral and central collisions, given that the impact parameter was integrated out in previous studies. The method for obtaining the impact parameter-dependent cross section was developed in the 1990s \cite{impact1,impact2}. This rigorous approach may yield results significantly different from those derived from the equivalent photon approximation, particularly in explaining the transverse momentum spectrum of lepton pairs at low transverse momentum \cite{lowTMD1,lowTMD2}. Additionally, variations in experimental methods for extracting UPC events can introduce potential errors since the impact parameters are not directly observable.

The paper is organized as follows. This section introduces our motivation and relevant background. The next section derives the cross section for the impact parameter dependence of the $\gamma \gamma \rightarrow l^{+} l^{-} \gamma$  process and provides numerical estimates for three different UPC cases within the kinematic ranges of RHIC and LHC. The last section offers a brief summary and outlook, while detailed derivations are included in the Appendix.

\section{Impact parameter-dependent cross sections of \texorpdfstring{$\gamma \gamma \rightarrow l^+ l^- \gamma$}{gamma gamma -> l+ l- gamma} in UPCs}


In UPCs, the primary interaction involves the collision of electromagnetic fields. The electromagnetic fields can be viewed as two quasi real beams of photons, which can then annihilate to produce dileptons or quark pair. Here, we primarily focus on the simple process of photon annihilation into dileptons.

\begin{equation}
\gamma_1 (x_1 P + k_{1\perp}) + \gamma_2 (x_2 \bar{P} + k_{2\perp}) \rightarrow l^+ (p_1) + l^- (p_2)
\end{equation}
When a high-energy heavy ion emits a photon, the dominant component of its momentum is longitudinal, making $k_{1\perp}$ and $k_{2\perp}$ relatively small. In this paper, we explore the impact of the background field on the coupling constant and investigate the process where the final-state dilepton emits a photon, then.

\begin{equation}
\gamma_1 (x_1 P + k_{1\perp}) + \gamma_2 (x_2 \bar{P} + k_{2\perp}) \rightarrow l^+ (p_1) + l^- (p_2) + \gamma(k)
\end{equation}
As the emission of a photon by leptons occurs within the peripheral electric field of the heavy ion and is thus influenced by the background electric field, which is the focus of this study.

Before deriving the impact parameter-dependent cross section, we need to address two ingredients. First, we must quantitatively assess how the background field affects the coupling constant for subsequent numerical estimates. Second, we need to parameterize the background field, which describes the distribution function of quasi-real photons outside the nucleus. In previous studies, we briefly examined the impact of the background field on the coupling constant in the same process. Below, we provide a brief overview of how these two ingredients were addressed in earlier work \cite{Li3}. To quantitatively assess the effect, we begin with the photon propagator in a photon background field,
\begin{equation}
    \begin{aligned}
 \left\langle B\left|A_\mu(x) A_v(y)\right| B\right\rangle& =\int \frac{d^3 k}{(2 \pi)^3 2 k_0} \int \frac{d^3 k^{\prime}}{(2 \pi)^3 2 k_0^{\prime}}\langle B \mid k\rangle\left\langle k\left|A_\mu(x) A_v(y)\right| k^{\prime}\right\rangle\left\langle k^{\prime} \mid B\right\rangle \\
& =\int \frac{d^3 k}{(2 \pi)^3 2 k_0}\langle B \mid k\rangle\left\langle k\left|A_\mu(x) A_v(y)\right| k\right\rangle\langle k \mid B\rangle  \\
& =\int \frac{d^3 k}{(2 \pi)^3 2 k_0}|\langle B \mid k\rangle|^2 \mathrm{e}^{-i k \cdot(x-\mathrm{y})} g_{\mu v} \\
& =\int \frac{d^3 k}{(2 \pi)^3 2 k_0}|\langle B \mid k\rangle|^2 \mathrm{e}^{-i k \cdot(x-\mathrm{y})} g_{\mu v}  \\
& =\int \frac{d^4 k}{(2 \pi)^4} \mathrm{e}^{-i k \cdot(x-\mathrm{y})} \frac{-\mathrm{i} g_{\mu v}}{k^2+\mathrm{i} \varepsilon}|\langle B \mid k\rangle|^2
\end{aligned}
\end{equation}
where $B$ represents the background photon field corresponding to the peripheral electric fields of the high-speed moving heavy ions in this work. The detailed derivation requires the Wick contraction. The above expression indicates that the photon propagator in the background field differs from that in a vacuum, being convolved with a background photon distribution function $|\langle B | k \rangle|^2 = f(k)$. This function represents the probability density distribution of real photons with momentum $k$ in a non-polarized background field. If we overlook the alteration of the propagator in the background field, we might neglect $f(k)$ or compensate by using an effective coupling constant,
\begin{equation}
\alpha_{\text{eff}}(k) = \alpha_e f(k)
\end{equation}
where $\alpha_e$ is the electromagnetic coupling constant (approximately $1/137$). $\alpha_{\text{eff}}$ thus represents the actual electromagnetic interaction strength measured in the background field. Eq. 4 describes how the background field modifies the coupling constant, which can be used to replace the coupling constant in the vertex associated with lepton radiation photons.

The second ingredient involves parameterizing the background electric field, specifically the form of $f(k)$ in Eq. 4. We typically employ the equivalent photon approximation to describe the electric field around heavy ions, treating the electromagnetic field as a beam of photons with a known distribution function. This method is effective and has been validated by numerous theoretical studies and experiments \cite{Li1,Li2,Shao1,star2021}. Under the equivalent photon approximation \cite{EPA1,EPA2}, the photon distribution function is given by,

\begin{equation}
n(k_z, k_\perp^2) = \frac{Z^2 \alpha_e}{\pi^2} \frac{k_\perp^2}{|k_z|} \left[ \frac{F(k_\perp^2 + k_z^2)}{(k_\perp^2 + k_z^2)} \right]^2,
\end{equation}
where $Z$ is the nuclear charge number and $F$ is the nuclear charge form factor. The relationship between $n$ here and $f$ in Eq. 4 is $n(|k|)=\frac{f(|k|)}{2k_0 (2\pi)^3}$. We can use a parameterized form of the momentum-space form factor from the STARlight MC generator \cite{light},

\begin{equation}
F(|\vec{k}|) = \frac{4\pi \rho^0}{|\vec{k}|^3 A} \left[\sin(|\vec{k}| R_A) - |\vec{k}| R_A \cos(|\vec{k}| R_A)\right] \frac{1}{(a^2 |\vec{k}|^2 + 1)}
\end{equation}
where $R_A$ is the nuclear radius, typically defined as $R_A = 1.1 A^{1/3} \, \text{fm}$ (for Pb and Au) and $R_A = 1.2 A^{1/3} \, \text{fm}$(for Ru), and $a = 0.7 \, \text{fm}$. $\rho^0$ is the normalization factor. This parameterized form closely approximates the Woods-Saxon distribution and will be frequently used in numerical evaluations.

Next, we refer to Refs. \cite{impact1,impact2} to incorporate the dependence of our cross section on the impact parameter. The process of the final lepton pair radiating a photon can be categorized into three cases.
\begin{figure}[ht]
	\centering 
	\includegraphics[width=0.7\textwidth, angle=0]{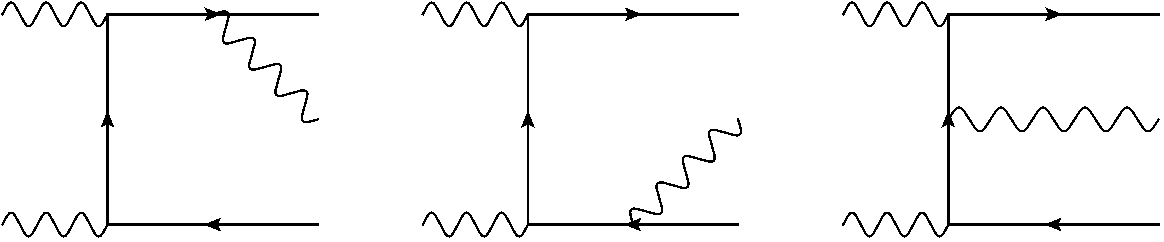}	
	\caption{The process $\gamma \gamma\rightarrow l^+ l^- \gamma$}
\end{figure}
Then, the cross section with dependence on the impact parameter is given by,

\begin{equation}
\begin{aligned}
& \frac{d \sigma}{d^2 p_{1 \perp} d^2 p_{2 \perp} d^2 k_{\perp} d y_1 d y_2 d y_3 d^2 b_{\perp}}=\frac{Z^4 \alpha_e^4}{Q^4}\left(4 \pi \alpha_{e f f}\right) \frac{4}{(2 \pi)^6}  \\
& \quad\quad\times \int d^2 k_{1 \perp} d^2 k_{2 \perp} d^2 \Delta_{\perp}\left[\mathcal{A}\left(k_{1 \perp} \cdot k_{1 \perp}^{\prime}\right)\left(k_{2 \perp} \cdot k_{2 \perp}^{\prime}\right)+\mathcal{B}\left(k_{1 \perp} \cdot k_{2 \perp}^{\prime}\right)\left(k_{2 \perp} \cdot k_{1 \perp}^{\prime}\right)+\mathcal{C}\left(k_{1 \perp} \cdot k_{2 \perp}\right)\left(k_{1 \perp}^{\prime} \cdot k_{2 \perp}^{\prime}\right)\right] \\
& \quad\quad\times \delta^2\left(p_{1 \perp}+p_{2 \perp}+k_{\perp}-k_{1 \perp}-k_{2 \perp}\right) e^{i \Delta_{\perp} \cdot b_{\perp}} \mathcal{F}\left(x_1, k_{1 \perp}^2\right) \mathcal{F}^*\left(x_1, k_{1 \perp}^{\prime 2}\right) \mathcal{F}\left(x_2, k_{2 \perp}^2\right) \mathcal{F}^*\left(x_2, k_{2 \perp}^{\prime2}\right)
\end{aligned}
\end{equation}
where $\Delta_\perp = k_{1\perp} - k_{1\perp}' = k_{2\perp}' - k_{2\perp}$. The radial momentum of the incoming photons can be determined using energy and momentum conservation,
\begin{equation}
\begin{aligned}
x_1 \sqrt{s} &= \sqrt{p_{1\perp}^2 + m^2} e^{y_1}+\sqrt{p_{2\perp}^2 + m^2} e^{y_2}+\sqrt{k_\perp^2 } e^{y_3}  \quad\text{and}\\
x_2\sqrt{s}  &= \sqrt{p_{1\perp}^2 + m^2} e^{-y_1}+\sqrt{p_{2\perp}^2 + m^2} e^{-y_2}+\sqrt{k_\perp^2 } e^{-y_3}
\end{aligned}
\end{equation}
where $m$ is the lepton mass, which is approximately 0 for the electron, and $s$ is the center mass energy. The nuclear charge form factor is incorporated as $\mathcal{F}(x, k_\perp^2) = \frac{F(k_{\perp}^2 + x^2 M_p^2)}{(k_\perp^2 + x^2 M_p^2)}$, with $M_p$ being the mass of the proton. The coefficients $\mathcal{A}$, $\mathcal{B}$, and $\mathcal{C}$ are Lorentz invariants composed of the dot product of particle momenta before and after the interaction. Their specific forms can be found in Appendix. Notably, it is emphasized that the coupling constant is affected by the background field, which is accounted for by replacing the coupling constant $\alpha_{\text{e}}$ with $\alpha_{\text{eff}}$ for real photon radiation.

In previous studies \cite{Li3}, we compared changes in the cross section before and after considering the background field. However, we will not revisit that comparison here. After incorporating the impact parameter dependence into the cross section, we will numerically evaluate three distinct cases: UPC, centrality 60\%-99.9\%, and tagged UPC. For the UPC case, the impact parameters are integrated over the regions $[2R_A, \infty]$, indicating that the two nuclei barely pass each other. Since the impact parameter is not directly observable, we often use the Glauber model to derive it indirectly \cite{glauber}. The principle is that a smaller impact parameter correlates with a greater variety of final-state particles. In the STAR experiment at RHIC, the pair production cross section associated with dual electromagnetic excitation was measured. Neutrons emitted by the fragmenting nuclei are detected at forward angles and used as triggers for UPC. Requiring that lepton pairs be produced simultaneously with the Coulomb breakup of the beam nuclei changes the impact parameter distribution in comparison to exclusive production. To align theoretical result more closely with experimental results, one can define the "tagged" UPC cross section as,

\begin{equation}
2\pi \int_{2R_A}^{\infty} b_\perp db_\perp P_\perp(b_\perp) d\sigma(b_\perp, ...),
\end{equation}
where $P_\perp$ is the probability of neutron emission from the scattered nucleus, parameterized as \cite{taggedUPC},
\begin{equation}
P(b_\perp) = 5.45 \times 10^{-5} \frac{Z^3 (A-Z)}{A^{2/3} b_\perp^2} e^{-5.45 \times 10^{-5} \frac{Z^3 (A-Z)}{A^{2/3} b_\perp^2}}
\end{equation}
In the context of Coulomb dissociation interactions, the mean impact parameter is significantly reduced.

Finally, we present the results of our numerical evaluation. For the process of photon annihilation into electron pairs accompanied by photon emission, we plot the dependence of the cross section on the final photon energy $|k|$ for Au-Au and Ru-Ru collisions at a center-of-mass energy of \(\sqrt{s} = 200 \, \text{GeV}\). The rapidity of the dilepton is restricted to the range $[-1, 0]$ and $p_{1\perp}, p_{2\perp} \in [0.2, 0.8] \, \text{GeV}$, as shown in Figures 2 and 3. The integration interval for the final photon polar angle is set to correspond to the rapidity interval $[0, 1]$ to avoid collinear divergence. In each figure, we evaluate the differential cross section for the three cases listed earlier. These plots show that the cross section's dependence on photon energy is a monotonically decreasing function that falls off rapidly. In the same figure, the relative magnitudes of the cross sections among the three different UPC cases remain consistent at the same photon energy, with the UPC case yielding the largest cross section and the tagged UPC case the smallest. This indicates a significant dependence on the impact parameter, that inconsistently selecting UPC cases can lead to substantial errors in both theoretical calculations and experimental analyses.
\begin{figure}[ht]
	\centering 
	\includegraphics[width=0.5\textwidth, angle=0]{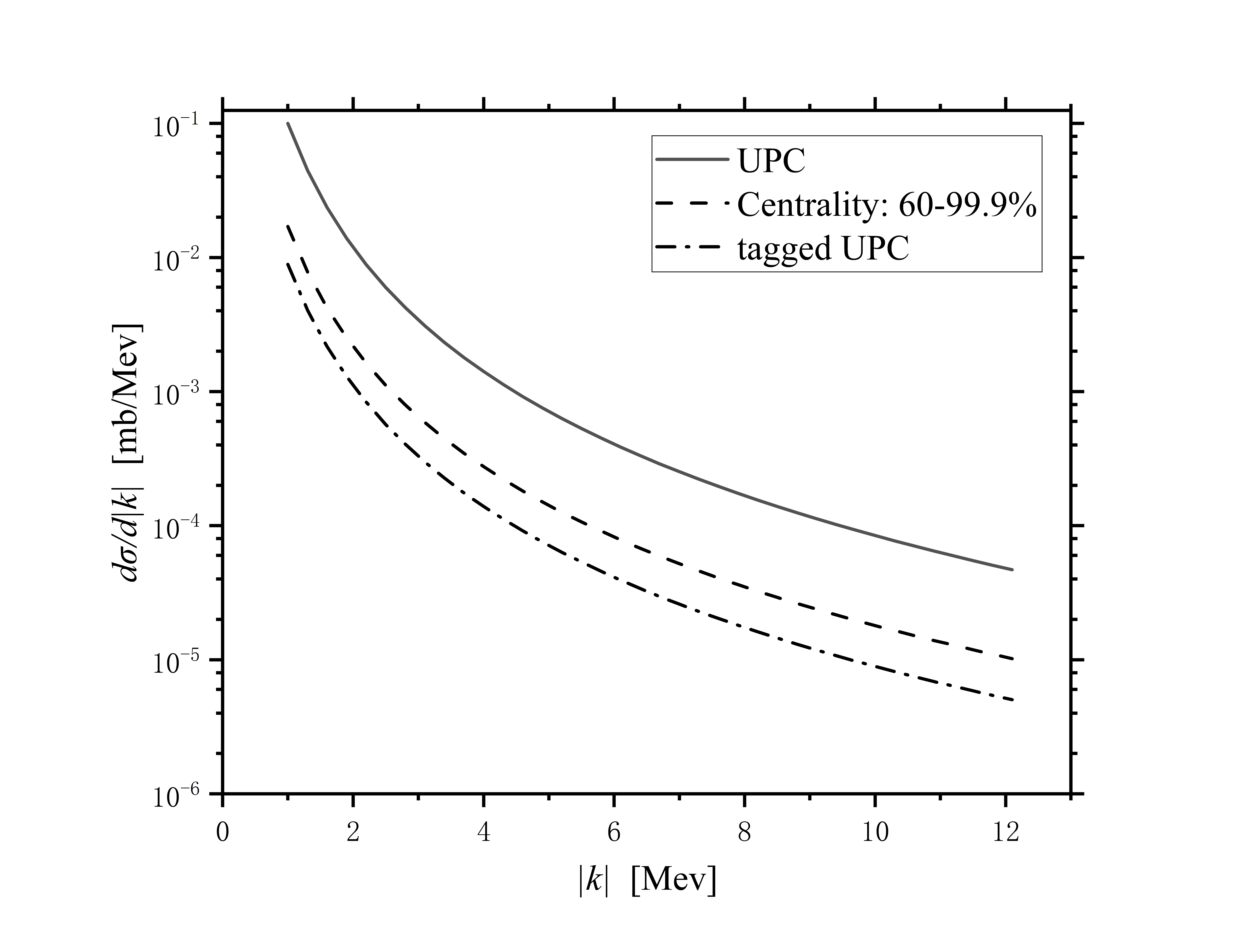}	
	\caption{Estimates of the cross section of Au-Au as the function of $|k|$ for the different centralities at $\sqrt{s}$=200GeV. The integration interval for the final state photon's rapidity is $[0, 1]$. The integration intervals for the electron pair's rapidity and transverse momentum are $[-1, 0]$ and $[0.2, 0.8]\, \text{GeV}$, respectively.}
\end{figure}
\begin{figure}[ht]
	\centering 
	\includegraphics[width=0.5\textwidth, angle=0]{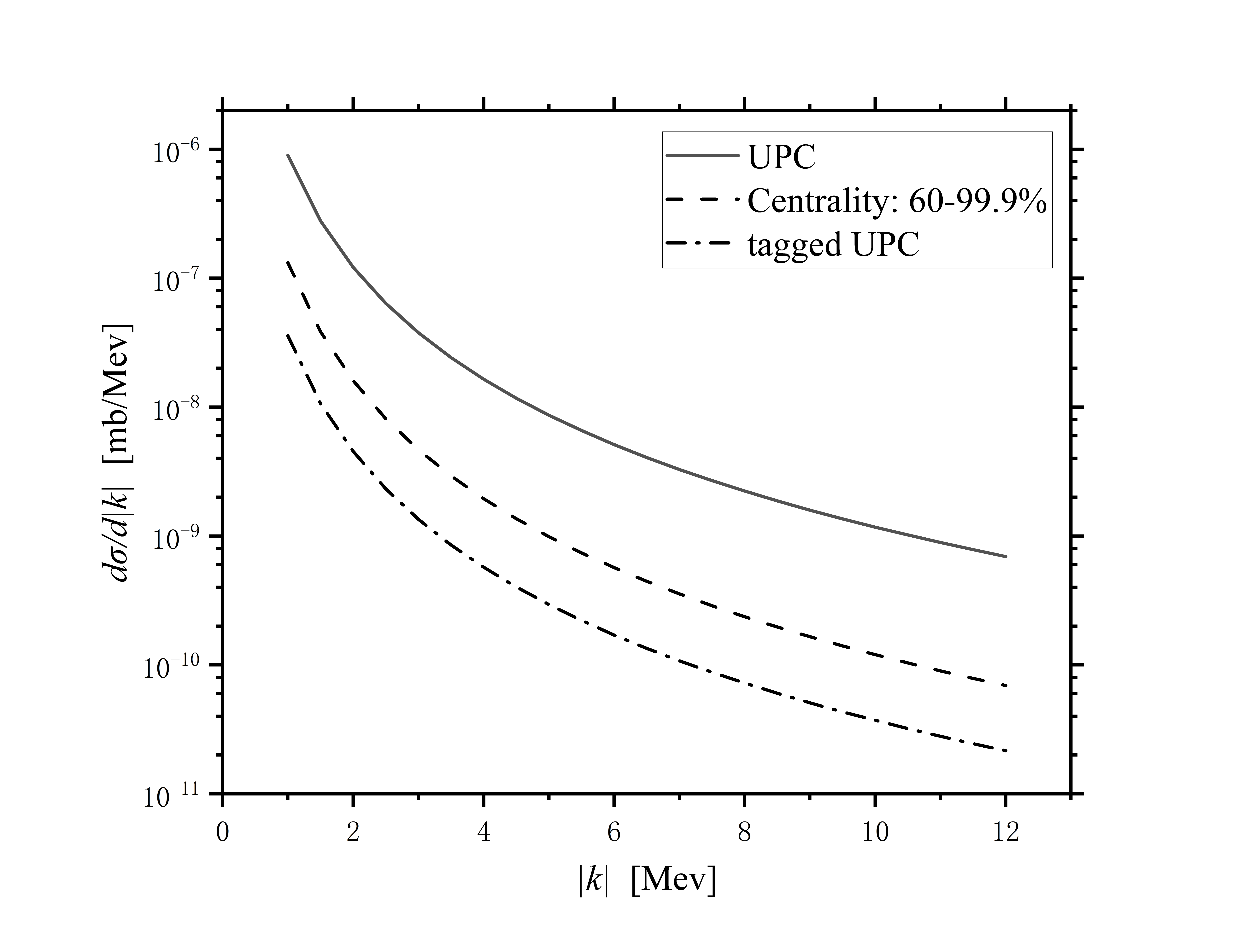}	
	\caption{Estimates of the cross section of Ru-Ru as the function of $|k|$ for the different centralities at $\sqrt{s}$=200GeV. The integration interval for the final state photon's rapidity is $[0, 1]$. The integration intervals for the electron pair's rapidity and transverse momentum are $[-1, 0]$ and $[0.2, 0.8]\, \text{GeV}$, respectively.}
\end{figure}
By comparing Figure 2 with Figure 3, we observe that, under the same center mass energy, the cross section for light nuclei is smaller. This result aligns with expectations for two reasons: first, the photon beam parameterized from heavy nuclei exhibits higher brightness, and second, the high-density background field produced by heavy nuclei leads to greater corrections to the coupling constant. This reinforces our central theme that a dense photon background field will have a larger impact on the coupling constant.

Additionally, we also evaluated the numerical results for Pb-Pb collisions. For the same process of photon annihilation into electron pairs accompanied by photon radiation, we present in Figure 4 the dependence of the cross section on the final state photon energy $|k|$ for three UPC cases at a center mass energy of $\sqrt{s} = 5.02 \, \text{TeV}$, with the dilepton rapidity constrained to the $[-1, 0]$ region and $p_{1\perp}, p_{2\perp} \in [0.2, 0.8] \, \text{GeV}$. Comparing Figures 2 and 4, we find that the dependence of this process on the center mass energy is not particularly pronounced. Furthermore, due to the similarity in the atomic numbers of Au(79)  and Pb(82), their cross section magnitudes are closely aligned.

\begin{figure}[ht]
	\centering 
	\includegraphics[width=0.5\textwidth, angle=0]{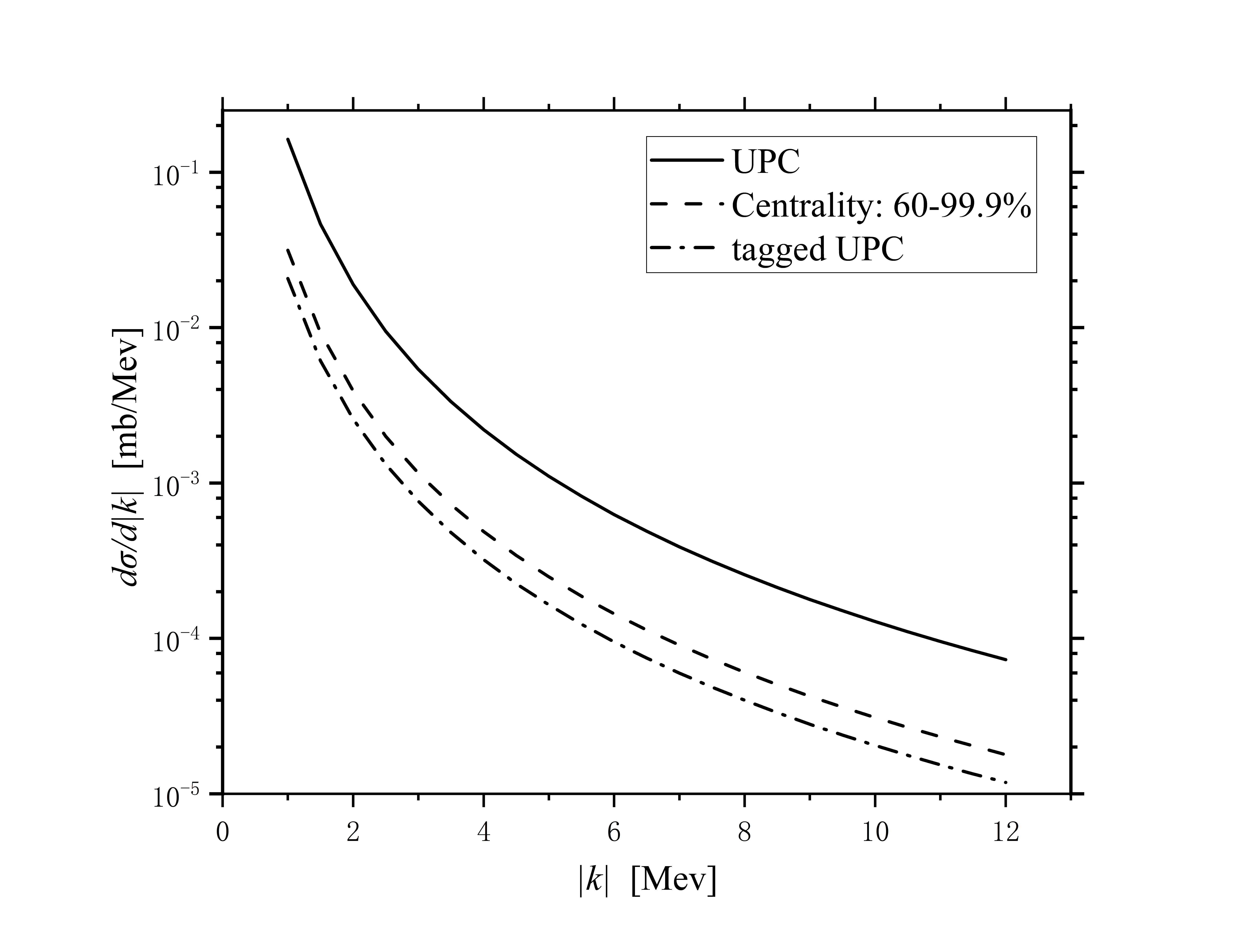}	
	\caption{Estimates of the cross section of Pb-Pb as the function of $|k|$ for the different centralities at $\sqrt{s}$=5.02TeV. The integration interval for the final state photon's rapidity is $[0, 1]$. The integration intervals for the electron pair's rapidity and transverse momentum are $[-1, 0]$ and $[0.2, 0.8]\, \text{GeV}$, respectively.}
\end{figure}

\section{Summary and Outlook}

In this paper, we study the effect of background fields on photons radiated from electrons. The primary motivation is that the vertex of photon radiation is influenced by the background field, which is the focus of our investigation. We selected the ultra-peripheral collision process, a popular area of research in recent years, where the extranuclear electric field of heavy ions can be parameterized as a beam of photons. This approach provides us with a stable and known background field for our study. To better distinguish between different UPC cases, we introduce a dependence on the impact parameter. It is important to note that there are three UPC cases, and the reason is that the impact parameter is not directly observable in experiments, necessitating various methods to deduce these parameters from experimental results. Finally, we estimate the cross section for three different cases within the kinematic range between RHIC and LHC.

From our numerical estimation results, it is evident that the cross section of this process decreases monotonically and rapidly with the energy of the final photon. Moreover, the results for different UPC cases vary significantly, indicating a strong dependence on the impact parameter. Careful treatment of the impact parameter in theoretical calculations and experimental analyses is essential, as it could be a major source of theoretical and experimental error. By comparing the center mass energy of the Au-Au collision at 200 GeV with that of the Pb-Pb collision at 5.02 TeV, we found that the impact of center mass energy on this process is not particularly pronounced. In contrast, comparing the same center mass energie of the Au-Au and Ru-Ru collisions at 200 GeV revealed that the nuclear charge number significantly influences this process. It is evident that the photons associated with the strong electric field generated by heavy nuclei are denser, resulting in a dense photon field that significantly affects the coupling constant and substantially enhances the interaction strength. This conclusion aligns with current experimental observations, where we have identified a non-perturbative problem arising from strong interactions in the kinematic range associated with perturbative QCD under a dense gluon background field.

There are some shortcomings and prospects to consider. The issue of soft photon radiation from final state charges has been discussed in some literature but is not addressed here \cite{sudakov}. Since it is experimentally challenging to distinguish whether the final state real electrons radiate soft photons, the theoretical results should encompass all possible soft photon radiation cases. However, compared to existing soft photon radiation corrections \cite{sudakov}, the soft photon radiation in a background field is more complex. This complexity arises because the corresponding coupling constant for soft photon radiation is modified by the background field, complicating the resummation process. Nevertheless, finite-order summation may help reduce the discrepancy with experimental results to some extent, easing the theoretical correction of the effect, which warrants further investigation.

\section*{Acknowledgments}
AI tools were used in this paper to improve sentence fluency and make the English more natural.

\appendix

\section{The specific forms of the coefficients \texorpdfstring{$\mathcal{A}$}{A}, \texorpdfstring{$\mathcal{B}$}{B} and \texorpdfstring{$\mathcal{C}$}{B} }

Referring to \cite{impact1,impact2}, we obtain the impact parameter-dependent cross section, which is given by

\begin{equation}
\begin{aligned}
&\frac{d \sigma}{d^2 p_{1 \perp} d^2 p_{2 \perp} d^2 k_{\perp} d y_1 d y_2 d y_3}=\frac{Z^4 \alpha_e^4}{Q^4}\left(4 \pi \alpha_{e f f}\right) \frac{4}{(2 \pi)^6}  \\
&\quad\quad\quad\quad \times \int d^2 k_{1 \perp} d^2 k_{2 \perp} d^2 \Delta_{\perp}\left[k_{1 \perp}^i k_{2 \perp}^j k_{1 \perp}^{\prime \mu} k_{2 \perp}^{\prime v} \Gamma_{i j}\left(k_1, k_2, p_1, p_2, k\right) \Gamma_{\mu \nu}^*\left(k_1, k_2, p_1, p_2, k\right)\right]_{k_1+k_2-p_1-p_2-k=0} \\
&\quad\quad\quad\quad \times \delta^2\left(p_{1 \perp}+p_{2 \perp}+k_{\perp}-k_{1 \perp}-k_{2 \perp}\right) e^{i \Delta_{\perp} . b_{\perp}} \mathcal{F}\left(x_1, k_{1 \perp}^2\right) \mathcal{F}^*\left(x_1, k_{1 \perp}^{\prime 2}\right) \mathcal{F}\left(x_2, k_{2 \perp}^2\right) \mathcal{F}^*\left(x_2, k_{2 \perp}^{\prime}{ }^2\right)
\end{aligned}
\end{equation}
where $\Gamma^{ij}(k_1, k_2, p_1, p_2, k)$ is the vertex function of the reaction process. After contraction, we have

\begin{equation}
\begin{aligned}
k_{1 \perp}^i k_{2 \perp}^j k_{1 \perp}^{\prime \mu} k_{2 \perp}^{\prime v} &\Gamma_{i j}\left(k_1, k_2, p_1, p_2, k\right) \Gamma_{\mu \nu}^*\left(k_1, k_2, p_1, p_2, k\right) \\
& \quad=\mathcal{A}\left(k_{1 \perp} \cdot k_{1 \perp}^{\prime}\right)\left(k_{2 \perp} \cdot k_{2 \perp}^{\prime}\right)+\mathcal{B}\left(k_{1 \perp} \cdot k_{2 \perp}^{\prime}\right)\left(k_{2 \perp} \cdot k_{1 \perp}^{\prime}\right)+\mathcal{C}\left(k_{1 \perp} \cdot k_{2 \perp}\right)\left(k_{1 \perp}^{\prime} \cdot k_{2 \perp}^{\prime}\right)
\end{aligned}
\end{equation}
and the coefficients $\mathcal{A}$, $\mathcal{B}$, and $\mathcal{C}$ are defined as:

\begin{equation}
\begin{aligned}
\mathcal{A}=&-\frac{1}{2 \alpha_1 \alpha_2 \beta_1\left(-\gamma+\alpha_1+\beta_1\right)^2 \beta_2\left(-\gamma+\alpha_2+\beta_2\right)^2} Q^2\left(2 \beta_2\left(2 \alpha_2^2+\left(2 \beta_2-3 \gamma\right) \alpha_2+2 \gamma\left(\gamma-\beta_2\right)\right) \alpha_1^3+\left(4 \beta_1 \alpha_2^3\right.\right. \\
& -2\left(5 \gamma \beta_1+3\left(\gamma-4 \beta_1\right) \beta_2\right) \alpha_2^2+\left(6 \beta_1\left(\gamma-4 \beta_2\right)\left(\gamma-\beta_2\right)+\gamma\left(\gamma Q^2-6 \beta_2^2+8 \gamma \beta_2\right)\right) \alpha_2-\left(\gamma-\beta_2\right)\left(Q^2 \gamma^2\right. \\
& \left.\left.+\left(Q^2+6 \gamma-10 \beta_1\right) \beta_2 \gamma+4 \beta_1 \beta_2^2\right)\right) \alpha_1^2+\left(2 \beta_1\left(2 \beta_1-\gamma\right) \alpha_2^3+\left(2\left(\gamma^2-9 \beta_1 \gamma+12 \beta_1^2\right) \beta_2-\gamma\left(\gamma Q^2\right.\right.\right. \\
& \left.\left.+14 \beta_1^2-8 \gamma \beta_1\right)\right) \alpha_2^2+\left(\beta_1\left(3 Q^2-6 \gamma+10 \beta_1\right) \gamma^2-\left(\gamma\left(Q^2+2 \gamma\right)+2 \beta_1\left(Q^2-10 \gamma+15 \beta_1\right)\right) \beta_2 \gamma+2\left(\gamma^2\right.\right. \\
& \left.\left.-9 \beta_1 \gamma+12 \beta_1^2\right) \beta_2^2\right) \alpha_2+\left(\gamma-\beta_2\right)\left(Q^2\left(\gamma-3 \beta_1\right) \gamma^2+2\left(\gamma\left(Q^2+\gamma\right)+3 \beta_1\left(\beta_1-\gamma\right)\right) \beta_2 \gamma+2(\gamma\right. \\
& \left.\left.\left.-2 \beta_1\right) \beta_1 \beta_2^2\right)\right) \alpha_1+\left(\gamma-\beta_1\right)\left(Q^2\left(\gamma-\beta_2\right)\left(\beta_1-\beta_2\right) \gamma^2+\alpha_2^2\left(Q^2 \gamma^2-\beta_1\left(Q^2+2 \gamma-2 \beta_2\right) \gamma+4 \beta_1^2(\gamma\right.\right. \\
& \left.\left.\left.\left.-\beta_2\right)\right)+\alpha_2\left(Q^2\left(\beta_2-\gamma\right) \gamma^2+2 \beta_1\left(\gamma^2-\beta_2 \gamma+\beta_2^2\right) \gamma+\beta_1^2\left(-4 \gamma^2+6 \beta_2 \gamma-4 \beta_2^2\right)\right)\right)\right) \\
\end{aligned}
\end{equation}
\begin{equation}
\begin{aligned}
\mathcal{B}=&\frac{1}{2 \alpha_1 \alpha_2 \beta_1\left(-\gamma+\alpha_1+\beta_1\right)^2 \beta_2\left(-\gamma+\alpha_2+\beta_2\right)^2} Q^2\left(2 \beta_2\left(\alpha_2\left(\gamma-2 \beta_2\right)+2\left(\gamma-\beta_2\right) \beta_2\right) \alpha_1^3+\left(4\left(2 \gamma-\beta_1\right) \beta_2^3\right.\right. \\
& +\left(\gamma \left(Q^2-10 \gamma+10 \alpha_2\right)+2\left(\gamma+2 \alpha_2\right) \beta_1\right) \beta_2^2+2\left(\left(\gamma+\beta_1\right) \gamma^2-\alpha_2\left(4 \gamma+3 \beta_1\right) \gamma+\alpha_2^2\left(\gamma+4 \beta_1\right)\right) \beta_2-\gamma(\gamma \\
& \left.\left.-\alpha_2\right)\left(Q^2 \gamma-2 \alpha_2 \beta_1\right)\right) \alpha_1^2+\left(2\left(3 \gamma-2 \beta_1\right) \beta_1 \alpha_2^3+\left(-\left(\left(Q^2+2 \beta_2\right) \gamma^2\right)+2 \beta_1\left(3 \beta_2-4 \gamma\right) \gamma+2 \beta_1^2(\gamma\right.\right. \\
& \left.\left.+2 \beta_2\right)\right) \alpha_2^2+\left(\beta_1\left(3 Q^2+2 \gamma+2 \beta_1\right) \gamma^2-\left(\left(Q^2-6 \gamma\right) \gamma+2 \beta_1\left(Q^2+6 \gamma+3 \beta_1\right)\right) \beta_2 \gamma+2\left(-3 \gamma^2+3 \beta_1 \gamma\right.\right. \\
& \left.\left.\left.+4 \beta_1^2\right) \beta_2^2\right) \alpha_2+\gamma\left(\gamma-\beta_2\right)\left(\gamma\left(\gamma-3 \beta_1\right) Q^2+2\left(2 \gamma-3 \beta_1\right) \beta_2^2+2\left(\left(Q^2-\gamma\right) \gamma+\beta_1\left(\gamma+\beta_1\right)\right) \beta_2\right)\right) \alpha_1+(\gamma \\
& \left.-\beta_1\right)\left(4 \beta_1\left(\beta_1-\gamma\right) \alpha_2^3+\left(Q^2 \gamma^2-\beta_1\left(4 \beta_1\left(\gamma-\beta_2\right)+\gamma\left(Q^2-6 \gamma+6 \beta_2\right)\right)\right) \alpha_2^2+\gamma\left(\gamma\left(\beta_2-\gamma\right) Q^2-2 \beta_1^2 \beta_2\right.\right. \\
& \left.\left.\left.-2 \beta_1\left(\gamma^2-3 \beta_2 \gamma+\beta_2^2\right)\right) \alpha_2+Q^2 \gamma^2\left(\gamma-\beta_2\right)\left(\beta_1-\beta_2\right)\right)\right) \\
\end{aligned}
\end{equation}
\begin{equation}
\begin{aligned}
\mathcal{C}=&\frac{1}{2 \alpha_1 \alpha_2 \beta_1\left(-\gamma+\alpha_1+\beta_1\right)^2 \beta_2\left(-\gamma+\alpha_2+\beta_2\right)^2}\left(8 \beta_2^2\left(\beta_2-\gamma\right) \alpha_1^4+2 \beta_2\left(8 \beta_1 \alpha_2^2+\left(\left(2 \beta_2-3 \gamma\right) Q^2+4(\gamma \right.\right.\right. \\
& \left.\left.\left.-\beta_2\right)\left(\beta_2-3 \beta_1\right)\right) \alpha_2+4\left(\gamma-\beta_2\right)\left(\gamma Q^2+\beta_2\left(\gamma-3 \beta_1+\beta_2\right)\right)\right) \alpha_1^3+\left(8\left(\gamma-2 \beta_1\right) \beta_2^4+4\left(2 \alpha_2\left(\gamma-5 \beta_1\right)\right.\right. \\
& \left.+\left(\gamma+\beta_1\right)\left(Q^2-2 \gamma+4 \beta_1\right)\right) \beta_2^3+\left(-40 \beta_1 \alpha_2^2+\left(8 \beta_1\left(2 Q^2+4 \gamma+5 \beta_1\right)-2 \gamma\left(Q^2+4 \gamma\right)\right) \alpha_2+\gamma\left(Q^4\right.\right. \\
& \left.\left.+6 \gamma Q^2-2 \beta_1\left(13 Q^2-4 \gamma+8 \beta_1\right)\right)\right) \beta_2^2+2\left(-8 \beta_1 \alpha_2^3+\left(\gamma Q^2+4 \beta_1\left(Q^2+2 \gamma+5 \beta_1\right)\right) \alpha_2^2+\gamma\left(2 \gamma Q^2\right.\right. \\
& \left.\left.\left.+\left(-15 Q^2+8 \gamma-24 \beta_1\right) \beta_1\right) \alpha_2+Q^2 \gamma^2\left(11 \beta_1-5 \gamma\right)\right) \beta_2-\left(\gamma-\alpha_2\right)\left(Q^2 \gamma-8 \alpha_2 \beta_1\right)\left(Q^2 \gamma-2 \alpha_2 \beta_1\right)\right) \alpha_1^2 \\
& +\left(-16 \beta_1^2 \alpha_2^4+2 \beta_1\left(3 \gamma Q^2+2 \beta_1\left(Q^2+2 \gamma+6 \beta_1\right)+4\left(\gamma-5 \beta_1\right) \beta_2\right) \alpha_2^3+\left(24\left(\beta_2-\gamma\right) \beta_1^3+\left(-26 \gamma Q^2\right.\right.\right. \\
& \left.\left.+8 \gamma^2+8\left(2 Q^2+4 \gamma-5 \beta_2\right) \beta_2\right) \beta_1^2+2 \gamma\left(2 \gamma Q^2+\beta_2\left(3 Q^2-8 \gamma+8 \beta_2\right)\right) \beta_1-Q^2 \gamma^2\left(Q^2+2 \beta_2\right)\right) \alpha_2^2 \\
& +\left(8 \beta_2\left(2 \beta_2-3 \gamma\right) \beta_1^3+2\left(11 Q^2 \gamma^2+\beta_2\left(\gamma\left(8 \gamma-15 Q^2\right)+4\left(Q^2+2 \gamma-2 \beta_2\right) \beta_2\right)\right) \beta_1^2+\gamma\left(Q ^ { 2 } \left(3 Q^2\right.\right.\right. \\
& \left.\left.-10 \gamma) \gamma-2\left(Q^2-4 \beta_2\right) \beta_2\left(Q^2-2 \gamma+\beta_2\right)\right) \beta_1-Q^2 \gamma^2 \beta_2\left(Q^2-2 \gamma+2 \beta_2\right)\right) \alpha_2+Q^2 \gamma\left(\gamma-\beta_2\right)\left(10 \beta_2 \beta_1^2\right. \\
& \left.\left.-\left(3 \gamma Q^2+6 \beta_2^2+10 \gamma \beta_2\right) \beta_1+\gamma\left(Q^2+2 \beta_2\right)\left(\gamma+2 \beta_2\right)\right)\right) \alpha_1+\left(\gamma-\beta_1\right)\left(Q^2 \gamma\left(Q^2 \gamma-2 \alpha_2 \beta_1\right) \beta_2^2-\left(\gamma^2(\gamma\right.\right. \\
& \left.\left.-\alpha_2\right) Q^4+\gamma\left(\gamma Q^2+2 \alpha_2\left(\alpha_2-\gamma\right)\right) \beta_1 Q^2+2 \alpha_2\left(2\left(Q^2+2 \gamma-2 \alpha_2\right) \alpha_2-3 Q^2 \gamma\right) \beta_1^2\right) \beta_2+(\gamma \\
& \left.\left.\left.-\alpha_2\right)\left(-\gamma^2 \alpha_2 Q^4+\gamma\left(\gamma Q^2+\alpha_2\left(Q^2+2 \gamma+4 \alpha_2\right)\right) \beta_1 Q^2+8 \alpha_2^2 \beta_1^3-8 \alpha_2\left(\gamma Q^2+\alpha_2^2\right) \beta_1^2\right)\right)\right)
\end{aligned}
\end{equation}
where $\alpha_1=k_1 \cdot p_1$, $\alpha_2=k_1 \cdot p_2$, $\beta_1=k_2 \cdot p_1$, $\beta_2=k_2 \cdot p_2 $ and $\gamma=p_1 \cdot p_2$. The expressions for 4-momenta of the particles are $k_1=(0,x_1 P,\textbf{0}_\perp)$, $k_2=(x_2 \bar{P},0,\textbf{0}_\perp)$, $p_1=(\sqrt{p_{1\perp}^2+m^2} e^{y_1},\sqrt{p_{1\perp}^2+m^2} e^{-y_1},\textbf{p}_{1\perp})$, $p_2=(\sqrt{p_{2\perp}^2+m^2} e^{y_2},\sqrt{p_{2\perp}^2+m^2} e^{-y_2},\textbf{p}_{2\perp})$ and $k=(\sqrt{k_{\perp}^2} e^{y_3},\sqrt{k_{\perp}^2} e^{-y_3},\textbf{k}_{\perp})$. The $\bar{P}=P=\sqrt{s}$ is the center mass energy.

\bibliography{ref}

\end{document}